\documentclass[npg]{copernicus}
\usepackage{amsmath}
\usepackage{graphicx}
\usepackage{ifthen}
\usepackage{amsfonts}
\usepackage{amssymb}
\usepackage{epsfig}
\usepackage{url}

\def\be{\begin{equation}}
\def\ee{\end{equation}}

\begin{document}
\title{Another Look at Climate Sensitivity}
\author[1]{Ilya Zaliapin}
\author[2]{Michael Ghil}

\affil[1]{Department of Mathematics and Statistics,
University of Nevada, Reno, USA.
E-mail: zal@unr.edu.}

\affil[2]{Geosciences Department and Laboratoire de M\'{e}t\'{e}orologie 
Dynamique (CNRS and IPSL), Ecole Normale Sup\'{e}rieure, Paris, FRANCE, 
and Department of Atmospheric \& Oceanic Sciences and Institute of Geophysics 
\& Planetary Physics, University of California, Los Angeles, USA. 
E-mail: ghil@atmos.ucla.edu.}

\runningtitle{Climate Sensitivity}

\runningauthor{I. Zaliapin and M. Ghil}

\correspondence{Ilya Zaliapin (zal@unr.edu)}

\received{}
\pubdiscuss{} 
\revised{}
\accepted{}
\published{}

\firstpage{1}

\maketitle

\begin{abstract}
We revisit a recent claim 
that the Earth's climate system is characterized by sensitive dependence
to parameters; in particular, that the system exhibits an asymmetric, 
large-amplitude response to normally distributed feedback forcing.
Such a response would imply irreducible uncertainty in climate change
predictions and thus have notable implications for climate science and 
climate-related policy making. 
We show that equilibrium climate sensitivity in all generality does 
not support such an intrinsic indeterminacy; the latter appears
only in essentially linear systems.
The main flaw in the analysis that led to this claim is inappropriate 
linearization of an intrinsically nonlinear model; there is no room 
for physical interpretations or policy conclusions based on this 
mathematical error. Sensitive dependence nonetheless does exist 
in the climate system, as well as in climate models --- albeit in a very
different sense from the one claimed in the linear work under scrutiny --- and
we illustrate it using 
a classical energy balance model (EBM) with nonlinear feedbacks.
EBMs exhibit two saddle-node bifurcations, more recently called ``tipping points," 
which give rise to three distinct steady-state climates, two of which are
stable. Such bistable behavior is, furthermore, 
supported by results from more realistic, nonequilibrium climate models.  
In a truly nonlinear setting, indeterminacy in the size of the response 
is observed only in the vicinity of tipping points. 
We show, in fact, that small disturbances cannot result in a large-amplitude 
response, unless the system is at or near such a point.  
We discuss briefly how the distance to the bifurcation may be related to
the strength of Earth's ice-albedo feedback.  
\end{abstract}


\section{Introduction and motivation}
\label{intro}

\subsection{Climate sensitivity and its implications}
\label{implications}
Systems with feedbacks are an efficient mathematical tool for modeling
a wide range of natural phenomena; Earth's climate is one of the most
prominent examples.
Stability and sensitivity of feedback models is, accordingly, a traditional
topic of theoretical climate studies \citep{Cess76,MG76,CK78,Sch85,Sch86,
Cess89}. \citet{RB07} (RB07 hereafter) have recently advocated existence 
of intrinsically large sensitivities in an equilibrium model with multiple feedbacks.
Specifically, they argued that a small, normally distributed feedback may
lead to large-magnitude, asymmetrically distributed values of the system's 
response.  

Such a property, if valid, would have serious implications for climate
dynamics \citep{AF08} and for modeling of complex systems in general
\citep{WF08}.
In this paper, we revisit the dynamical behavior of a general, equilibrium climate model 
with genuinely nonlinear feedbacks, and focus subsequently on
a simple energy-balance model (EBM).
We notice that the main,  technical part of RB07's argument is well-known
in the climate literature, cf. \cite{Sch85,Sch86}, and thus it seems
useful to review the associated assumptions and possible interpretations 
of this result.

We rederive below in Section~\ref{RB} the key equation of RB07 and comment
on their purportedly nonlinear analysis in Section~\ref{RBNL}. 
We then proceed in 
Section \ref{sensitivity} with a more self-consistent version of sensitivity 
analysis for a nonlinear model. 
This analysis is applied in Section \ref{ebm} to a zero-dimensional EBM. 
Concluding remarks follow in Section \ref{discuss}.

\subsection{Roe and Baker's (2007) linear analysis}
\label{RB}
We follow here RB07 and assume the following general setup.
Let the net radiation $R$ at the top of the atmosphere be related to
the corresponding average temperature $T$ at the Earth's surface by
$R=R(T)$.
Assume, furthermore, that there exists a feedback $\alpha=\alpha(T)$, 
which is affected by the temperature change and which can, in turn, affect the 
radiative balance.
Hence, one can write $R=R\left(T,\alpha(T)\right)$.

To study how a small change $\Delta\,R$ in the radiation is related to the 
corresponding temperature change $\Delta\,T$, one can use the Taylor expansion 
\citep{Arf85} to obtain, as $\Delta\,T$ tends to zero,
\be
\Delta\,R=\frac{\partial\,R}{\partial\,T}\Delta\,T+
\frac{\partial\,R}{\partial\,\alpha}\frac{\partial\,\alpha}{\partial\,T}\Delta\,T
+\mathcal{O}\left((\Delta\,T)^2\right).
\label{Taylor}
\ee
Here, $\mathcal{O}(x)$ is a function such that $\mathcal{O}(x)\le C\,x$ as 
soon as $0<x<\epsilon$ for some positive constants $C$ and $\epsilon$.

Introducing the notations
\[\frac{1}{\lambda_0}=\frac{\partial\,R}{\partial\,T},~{\rm and~}
f=-\lambda_0\,\frac{\partial\,R}{\partial\,\alpha}\frac{\partial\,\alpha}{\partial\,T},\]
for the ``reference sensitivity'' $\lambda_0$ and the ``feedback factor'' $f$, we obtain
\be
\Delta\,R=\frac{1-f}{\lambda_0}\,\Delta\,T+\mathcal{O}\left((\Delta\,T)^2\right),
\label{eq0}
\ee
which readily leads to
\be
\Delta\,T=\frac{\lambda_0}{1-f}\Delta\,R 
+\mathcal{O}\left((\Delta\,T)^2\right),
\label{eq}
\ee
as long as $f\ne 1$. 

\begin{figure}[p]
\centering\includegraphics[width=8.3cm]{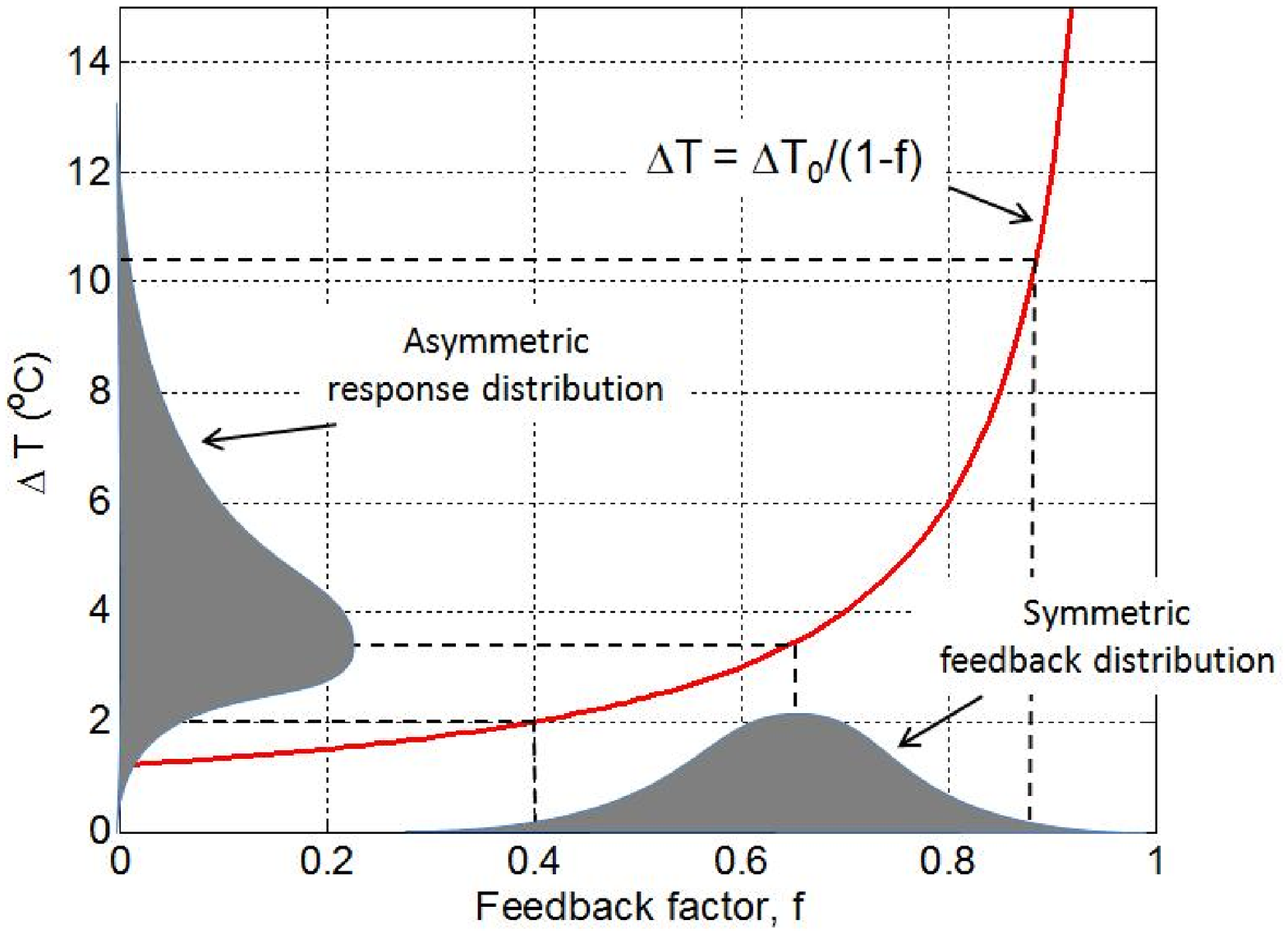}
\caption{Temperature response $\Delta\,T$ as
a function of the feedback factor $f$ in a linear climate model;
modified after \citet{RB07}.
In the absence of feedbacks ($f=0$), the response is given by 
$\Delta\,T=\Delta\,T_0=1.2^o$C. 
The response is amplified by feedbacks; it diverges 
($\Delta\,T\to \infty$) as $f\to1$.
The shaded areas illustrate the hypothetical symmetric distribution 
of the feedback factor $f$ and the corresponding asymmetric distribution
of the system response $\Delta\,T$.
Our study shows that this prominently asymmetric response is
only seen in a linear model and is absent in a general nonlinear
model (see Fig.~\ref{fig_RB_mod} below).
} 
\label{fig_RB}
\end{figure}

RB07 drop the higher-order terms in \eqref{eq} to obtain 
\be
\Delta\,T=\frac{\lambda_0}{1-f}\Delta\,R,
\label{eq1}
\ee
which is their equation (S4).
This equation leads directly to their main conclusion, namely that
a normally distributed feedback factor $f$ results
in an asymmetric system response $\Delta\,T$ to a fixed forcing $\Delta\,R$.
The purported sensitivity is due to the divergence of the right-hand side (rhs) 
of Eq.~(\ref{eq1}) [their equation (S4)] as $f$ approaches unity.
Figure~\ref{fig_RB}, which is analogous to Fig. 1 of RB07, 
illustrates this effect.

Roughly speaking, RB07 use the following argument: If the
derivative of $R(T)$ with respect to $T$ is close to 0, then the derivative 
of $T$ with respect to $R$ is very large, and a small change in the radiation 
$R$ corresponds to a large change in the temperature $T$.
Such an argument, though, is only valid for an essentially linear 
dependence $R\propto T$. Our straightforward analysis in 
Section~\ref{sensitivity} below shows that the sensitivity effect of Fig. 1 
is absent in climate models in which genuinely 
nonlinear feedbacks $R=R(T,\alpha(T))$ are present. 

It is worth noticing that, since one seeks the temperature change $\Delta\,T$ that results
from a change $\Delta\,R$ in the forcing, it might be preferable to consider the
inverse function $T=T(R)$ or, more precisely,
$T=T(R,\alpha(R))$ and the corresponding Taylor expansion
\[\Delta\,T=\frac{\partial\,T}{\partial\,R}\Delta\,R
+\frac{\partial\,T}{\partial\,\alpha}\frac{\partial\,\alpha}{\partial\,R}\Delta\,R
+\mathcal{O}\left((\Delta\,R)^2\right).\]
The main conclusions of our
analysis will not be affected by the particular choice of direct or inverse expansion, 
provided the nonlinearities are correctly taken into account.

\subsection{Roe and Baker's ``nonlinear" analysis}
\label{RBNL}
In this section we address the analysis carried out by RB07
in the supplemental on-line materials, pp. 4-5, Section ``Nonlinear feedbacks.''
The main conclusion of that analysis was that, given realistic
parameter values for the climate system, the effects of possible nonlinearities
in the behavior of the function $R=R(T)$ are negligible and do not affect
the system's sensitivity. We point out here two serious flaws in their mathematical
reasoning that, each separately and the two together, invalidate such a conclusion.

First, and most importantly, despite their section's title, the analysis carried 
out by RB07 is still {\it linear}. 
Indeed, the Taylor expansion in their Eq. (S7) is given by
\be
\Delta\,R\approx R'\Delta\,T+\frac{1}{2}\,R''\,\Delta\,T^2,
\label{S7}
\ee
where $(\cdot)'$ stands for differentiation with respect to $T$. But RB07 immediately 
invert this equation for $\Delta\,T$ subject to the assumption
\[\Delta\,T^2 = \Delta\,T\,\Delta\,T_0,\]
where $\Delta\,T_0$ is a constant.
Hence, instead of solving the quadratic Eq.~\eqref{S7}, Roe and Baker solve
the following {\it linear} approximation:
\[\Delta\,R = (R'+\frac{1}{2}\,R''\,\Delta\,T_0)\Delta\,T,\]
and thus obtain the key formula [their Eq. (S8)]
\be
\Delta\,T\approx\frac{-\lambda_0\,\Delta\,R}{1-f-\frac{1}{2}\lambda_0\,R''\,\Delta\,T_0};
\label{S8}
\ee
the last step uses the, correct, fact that $R' = (1-f)/\lambda_0$.
This equation artificially introduces a divergence point for the temperature at
$f = 1-(\lambda_0/2)\,R''\Delta\,T_0$, which clearly cannot exist
in a quadratic equation. 
Equation~\eqref{S8} is thus a very crude approximation that significantly 
deviates from the true solution to the full quadratic equation \eqref{S7} --- 
which we discuss below in Section~\ref{sensitivity} ---
and thus cannot be used to justify general statements about climate models.

The second flaw in the \citet{RB07} reasoning is that, using the model and 
parameter values they suggest, one readily finds that:
\begin{itemize}
\item $R' = -2$, {\it i.e.}, global temperature and radiation are 
{\it negatively} correlated, which is hardly the case for the current climate 
[{\it e.g.}, \cite{HeldSod}]. 
We notice that the negative sign of the correlation follows directly from 
their Eq.~(S10) and is not affected by particular values of the model parameters. 
Furthermore,
\item $R'' = -0.03$, which means that the model they consider is, indeed, 
essentially {\it linear}, and thus not very realistic.
\end{itemize}
Although, in this part of their analysis, Roe and Baker assumed that $f=0.4$, 
it is easy to check that, for all $f<0.95$, their model satisfies $|R''/R'|<0.1$
and is therewith very close to being linear. 
To conclude, the effects on nonlinearities are indeed negligible in the particular
model studied in this part of the RB07 paper, since the model is very close to
being linear; one cannot extrapolate, therefore, their conclusions to
climate models with significant nonlinearities. 

We next proceed with a mathematically correct sensitivity analysis of a general 
climate model in the presence of truly nonlinear feedbacks.

\section{A self-consistent sensitivity analysis}
\label{sensitivity}
It is easily seen from the discussion in Section \ref{RB}, especially from Eq.~\eqref{eq0}, 
that the relationship \eqref{eq1} is a crude approximation: it is valid only 
subject to the assumptions that ($a$) the higher-order terms in the 
expansion of $\Delta\,T$ are vanishingly small: 
\[\mathcal{O}\left((\Delta\,T)^2\right)\ll\frac{1-f}{\lambda_0}\Delta\,T;\]
and ($b$) the quantity in the rhs of this inequality is itself nonzero.

If one assumes, for instance, that 
$\mathcal{O}\left((\Delta\,T)^2\right)\sim C\,(\Delta\,T)^2$,
where the precise meaning of $g(x)\sim\,f(x)$ is given by 
$\lim_{x\to 0} g(x)/f(x) = 1$,
then the assumptions ({\it a,b}) above hold for $\Delta\,T$ that satisfy both
of the following conditions
\be
0<\Delta\,T\ll (1-f)/(\lambda_0\,C)\quad{\rm and}\quad
0<\Delta\,T< \epsilon,
\ee
where $C$ and $\epsilon$ are defined after Eq.~\eqref{Taylor}.
The first of these conditions implies that the range of temperatures 
within which the approximation \eqref{eq1} works vanishes as the feedback 
factor $f$ approaches unity. 
Hence, all the results based on this approximation --- including
precisely the main conclusions of RB07 --- no longer apply
outside a vanishingly small neighborhood of $f=1.0$. 

The asymptotic behavior we assumed above for 
$\mathcal{O}\left((\Delta\,T)^2\right)$ is not exotic.
Consider for instance the function $R=T^2$ in the neighborhood of $R=0$.
Its Taylor expansion
\[\Delta\,R = 2\,T\Delta\,T + \mathcal{O}\left((\Delta\,T)^2\right)\,\]
can be used to obtain, ignoring the second-order term,
\be
\Delta\,T \approx \Delta\,R/(2\,T).\
\label{contra}
\ee
The last equation would seem to imply that the growth of $\Delta\,T$ is inversely 
proportional to $T$ itself, so the change in $T$ should increase infinitely fast
as $T$ goes to 0, a rather annoying contradiction. 

The way out of this conundrum is to realize that 
the change $\Delta\,T$ given by Eq.~\eqref{contra}
is only valid in a small vicinity of $T=0$ and cannot be extrapolated to
larger values.
Of course, we all know that the function $R=T^2$ is nicely bounded and 
smooth in the vicinity of 0, but it is essential to take into account the
second term in its Taylor expansion in this vicinity to obtain correct
results.
We show in Section~\ref{ebm} below that this simple example
depicts the essential dependence of the Earth surface temperature 
on the global solar radiative input, for conditions close to those of
the current Earth system.

In summary, the linear approximation of the function $R(T)$ derived 
by RB07 from its Taylor expansion is not valid when $f$ approaches unity.
In this case --- which is precisely the situation emphasized by these 
authors --- the higher-order terms 
``hidden" inside $\mathcal{O}\left((\Delta\,T)^2\right)$, which they
neglected, are indispensable for a correct, self-consistent 
climate sensitivity analysis.

\begin{figure}[p]
\centering\includegraphics[width=8.3cm]{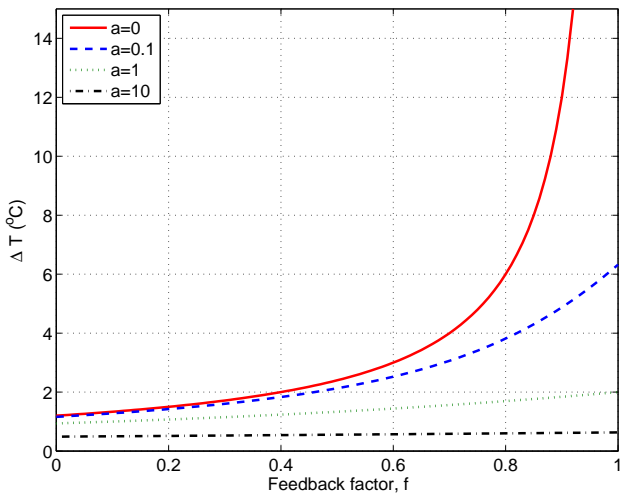}
\caption{Temperature response $\Delta\,T$ as
a function of feedback factor $f$ in a nonlinear, quadratic climate model,
governed by Eq.~\eqref{quadro}. The curves correspond (from top to bottom)  
to increasing values of the magnitude $a = R''/2$ 
of the quadratic term; the values of $a$ for each curve are given
in the figure's legend, in the upper-left corner.
The uppermost curve (red) corresponds to $a=0$
and is the same as the one shown in Fig.~\ref{fig_RB} above.
Extreme sensitivity, expressed in the divergence of $\Delta\,T$,
is only seen in the linear model; it rapidly vanishes in the
nonlinear model, as the nonlinearity factor $a$ increases.} 
\label{fig_RB_mod}
\end{figure}

A correct analysis of the case when $f$ approaches unity needs 
to start with a Taylor expansion that keeps the second-order term
\[\Delta\,R=\frac{1-f}{\lambda_0}\Delta\,T+a\,(\Delta\,T)^2+
\mathcal{O}\left((\Delta\,T)^3\right),\]
where $a=R''/2$.
If $\mathcal{O}\left((\Delta\,T)^3\right)$ is much smaller than
the other two terms on the rhs, then the temperature change
can be approximated by a solution of the quadratic equation
\be
\frac{1-f}{\lambda_0}\Delta\,T+a\,(\Delta\,T)^2=\Delta\,R.
\label{quadro}
\ee
The real-valued solutions to the latter equation, if they exist, are given by
\[\Delta\,T_{1,2}=\frac{1}{2}
\left(\frac{f-1}{a\,\lambda_0}\pm
\sqrt{\left(\frac{1-f}{a\,\lambda_0}\right)^2+\frac{4\,\Delta\,R}{a}}\right).\]
In particular, when $f$ is close to 1.0, then
\be
\Delta\,T_{1,2}\approx\pm
\sqrt{\frac{\Delta\,R}{a}}.
\label{TR}
\ee
One can see from Eq.~\eqref{TR} that the proximity of the feedback 
factor $f$ to unity no longer plays an important role in the qualitative 
behavior of the equilibrium temperature.
This point is further illustrated in Fig.~\ref{fig_RB_mod} that shows
the climate system's response $\Delta T$ as a function of the feedback 
factor $f$ for different values of the nonlinearity parameter $a$.
The most important observation is that the climate response does not diverge
at $f=1$; moreover, the asymmetry of the response due to the changes in
feedback factor $f$ rapidly vanishes as soon as the dependence of $\Delta\,R$ 
on $\Delta\,T$ becomes nonlinear.  

In general, one can consider an arbitrary number of terms in the Taylor
expansion of $R(T)$.
The very fact that one relies on the validity of the Taylor expansion 
implies that $R(T)$ is bounded and sufficiently smooth; in other words, 
a divergence of the equilibrium temperature due to a small change
in the forcing contradicts the very assumptions on which 
RB07 based their sensitivity analysis.

\section{Sensitivity for energy balance models (EBMs)}
\label{ebm}

We consider here a classical climate model with nonlinear feedbacks 
to illustrate that, in such a model: (i) the type of sensitivity 
claimed by RB07 does not exist; and (ii) sensitive dependence may 
exist, in a very different sense, namely in the neighborhood of bifurcation
points, as explained below.

\subsection{Model formulation}
\label{model}

We consider here a highly idealized type of model that connects the
Earth's temperature field to the solar radiative flux.
The key idea on which these models are built is due to 
\cite{Budyko69} and \cite{Sellers69}. They have been subsequently 
generalized and used for many studies of
climate stability and sensitivity; see \cite{HS74,GN75} and \cite{MG76}, 
among others.

The interest and usefulness of these ``toy" models resides in 
two complementary features: (i) their simplicity, which allows a complete
and thorough understanding of the key mechanisms involved;
and (ii) the fact that their conclusions have been extensively confirmed 
by studies using much more detailed and presumably realistic
models, including general circulation models (GCMs); see,
for instance, the reviews of \citet{NCC81} and \cite{GC87}.

The main assumption of EBMs is that the rate of change of the global 
average temperature $T$ is determined only by the net balance between
the absorbed radiation $R_{\rm i}$ and emitted radiation 
$R_{\rm o}$: 
\begin{equation}
c\frac{\mathrm{d}T}{\mathrm{d}t}= R_{\rm i}(T) 
- R_{\rm o}(T).
\label{T}
\end{equation} 

For simplicity, we follow here the  zero-dimensional (0-D) EBM
version of \cite{CK78} and \citet{GC87}, in which only global,
coordinate-independent quantities enter; thus
\be
\label{Ri}
R_{\rm i}=\mu\,Q_0\left(1-\alpha(T)\right),\quad
R_{\rm o}=\sigma\,g(T)\,T\,^4.
\ee
In the present formulation, the planetary {\it ice-albedo feedback} 
$\alpha$ decreases in an approximately linear fashion
with $T$, within an intermediate range of temperatures,
and is nearly constant for large and small $T$. Here
$Q_0$ is the reference value of the global mean solar radiative input,
$\sigma$ is the Stefan-Boltzmann constant, and
$g(T)$ is the grayness of the Earth, {\it i.e.} its deviation
from black-body radiation $\sigma\,T^4$.
The parameter $\mu\approx 1.0$ multiplying $Q_0$
indicates by how much the global insolation deviates
from its reference value.

We model the ice-albedo feedback by
\begin{equation}
\alpha(T;\kappa)=c_1+c_2\,
\frac{1-\tanh\left(\kappa\left(T-T_{\rm c}\right)\right)}{2}.
\label{albedo}
\end{equation}
This parametrization represents a smooth interpolation
between the piecewise-linear formula of Sellers-type models,
like those of \cite{MG76} or \cite{CK78}, and the 
piecewise-constant formula of Budyko-type models, 
like those of \cite{HS74} or \cite{GN75}.

Figure~\ref{fig_kappa}a shows four profiles of our
ice-albedo feedback $\alpha(T)=\alpha(T)$ as a function of $T$,
depending on the value of the steepness parameter $\kappa$.
The profile for $\kappa\gg 1$ would correspond roughly to
a Budyko-type model, in which the albedo $\alpha$ takes only 
two constant values, high and low, depending on whether
$T<T_{\rm c}$ or $T>T_{\rm c}$. The other profiles
shown in the figure for smaller values of $\kappa$, correspond 
to Sellers-type models, in which there exists a transition ramp 
between the high and low albedo values. 
Figure~\ref{fig_kappa}b shows the corresponding shapes of
the radiative input $R_{\rm i}=R_{\rm i}(T)$.

The {\it greenhouse effect} is parametrized, as in \cite{CK78} 
and \citet{GC87}, by letting
\be
\label{gh}
g(T)=1-m\,\tanh\left((T/T_0)^6\right).
\ee
Substituting this greenhouse effect parametrization and the
one for the albedo into Eq.~\eqref{T} leads to the following 
EBM:
\begin{eqnarray}
c\,\dot{T}=&\mu&\,Q_0\left(1-\alpha(T)\right)\nonumber\\
&-&\sigma\,T\,^4\left[1-m\,\tanh\left((T/T_0)^6\right)\right],
\label{T1}
\end{eqnarray}
where $\dot T=\mathrm{d}T/\mathrm{d}t$ denotes the 
derivative of global temperature $T$ with respect to time $t$.

It is important to note that current concern, both scientific and
public, is mostly with the greenhouse effect, rather than with
actual changes in insolation. But in a simple EBM model ---
whether globally averaged, like in \cite{CK78} and here, or 
coordinate-dependent,
as in \cite{Budyko69,Sellers69,HS74,GN75} or \cite{MG76}
--- increasing $\mu$ always results
in a net increase in the radiation balance. It is thus convenient,
and quite sufficient for the purpose at hand, to vary $\mu$
in the incoming radiation $R_{\rm i}$, rather than some other
parameter in the outgoing radiation $R_{\rm o}$. We shall 
return to this point in Section~\ref{discuss}.

\subsection{Model parameters}
The value $S$ of the {\em solar constant}, which is the value of the solar flux
normally incident at the top of the atmosphere along a straight line
connecting the Earth and the Sun, is assumed here to be $S=1370$ Wm$^{-1}$.
The reference value of the global mean solar radiative input is
$Q_0=S/4=342.5$, with the factor 1/4 due to Earth's sphericity.

The parameterization of the ice-albedo feedback in Eq.~\eqref{albedo}
assumes $T_c=273$ K and $c_1=0.15$, $c_2=0.7$, which ensures
that $\alpha(T)$ is bounded between 0.15 and 0.85, as in \cite{MG76};
see Fig.~\ref{fig_kappa}a. The greenhouse effect parametrization in 
Eq.~\eqref{gh} uses $m=0.4$, which corresponds to 
40\% cloud cover, and $T_0^{-6}=1.9\times10^{-15}$~K$^{-6}$ \citep{Sellers69,MG76}.
The Stefan-Boltzmann constant is $\sigma\approx5.6697\times10^{-8}$Wm$^{-2}$K$^{-4}$.

\subsection{Sensitivity and bifurcation analysis}
\subsubsection{Two types of sensitivity analysis}
We distinguish here between two types of sensitivity analysis for 
the 0-D EBM \eqref{T}.
In the first type, we assume that the system is driven 
out of an equilibrium state $T=T_0$, for which 
$R_{\rm i}(T_0) - R_{\rm o}(T_0)=0$, by an external force, 
and want to see whether and how it 
will return to a new equilibrium state, which may be different from 
the original one.
This analysis refers to the ``fast'' dynamics of the system, and 
assumes that 
$R_{\rm i}(T) - R_{\rm o}(T)\ne 0$ for $T\ne T_0$; it is often referred to 
as {\em linear stability analysis}, since it considers mainly small displacements
from equilibrium at $t=0$, $T(0)=T_0+ \theta(0)$, where $\theta(0)$ is of order 
$\epsilon$, with $0<\epsilon \ll 1$, as defined in Section \ref{sensitivity}.

The second type of analysis refers to the system's ``slow'' dynamics.
We are interested in how the system evolves along a {\it branch of 
equilibrium solutions} as the external force changes sufficiently slowly
for the system to track an equilibrium state;  hence, this second type of
analysis always assumes that the solution is in equilibrium with the
forcing: $R_{\rm i}(T) - R_{\rm o}(T)=0$ for all $T$ of interest.
Typically, we want to know how sensitive model solutions are
to such a slow change in a given parameter, and so this
type of analysis is called {\it sensitivity analysis}.
In the problem at hand, we will study --- again following \citet{CK78} 
and \citet{GC87} --- how changes in $\mu$, and hence in the global 
insolation, affect the model's equilibria.

A remarkable property of the EBM governed by Eq.~\eqref{T} 
is the existence of several stationary solutions  
that describe equilibrium climates of the Earth \citep{GC87}.
The existence and linear stability of these solutions  
result from a straightforward {\it bifurcation} analysis of 
the 0-D EBM (\ref{T}), as well as of its one-dimensional,
latutude-dependent counterparts
\citep{MG76,MG94}: there are two linearly 
stable solutions --- one that corresponds to the present
climate and one that corresponds to a much colder,
``snowball Earth'' \citep{HKHS98} --- separated by an 
unstable one, which lies about 10 K below the present climate.

The existence of the three equilibria --- two stable and
one unstable --- has been confirmed by such results being
obtained by several distinct EBMs, of either Budyko-
or Sellers-type \citep{NCC81,MG94}. Nonlinear stability, to large
perturbations in the initial state, has been investigated by
introducing a variational principle for the latitude-dependent 
EBMs of Sellers~\citep{MG76} and of Budyko~\citep{NCC81} 
type, and it confirms the linear stability results. 

\subsubsection{Sensitivity analysis for a 0-D EBM}

We analyze here the stability of the ``slow,'' quasi-adiabatic (in the
statistical-physics sense) dynamics of model \eqref{T1}. 
The energy-balance condition for steady-state solutions 
$R_{\rm i}=R_{\rm o}$ takes the form
\be
\mu\,Q_0\left(1-\alpha(T)\right)=
\sigma\,T\,^4\left[1-m\,\tanh\left((T/T_0)^6\right)\right].
\label{eql}
\ee
We assume here, following the previously cited EBM work, that the main 
bifurcation parameter is $\mu$; this happens to agree with the emphasis 
of \cite{RB07} on climate sensitivity as the dependence of mean temperature
$T$ on global solar radiative input, denoted here by $Q=\mu Q_0$. 

Figure~\ref{fig_TR} shows the absorbed and emitted radiative fluxes, 
$R_{\rm i}$ and $R_{\rm o}$, as functions of temperature $T$ for 
$\mu=0.5,1$ and 2.0.
One can see that Eq.~\eqref{eql} may have one or three 
solutions depending on the value of $\mu$: only the present, relatively
warm climate for $\mu=2.0$, only the ``deep-freeze" climate for $\mu=0.5$,
and all three, including the intermediate, unstable one for present-day
insolation values, $\mu=1.0$. 
These steady-state climate values are shown as a function of the 
insolation parameter $\mu$ in the bifurcation diagram of Fig.~\ref{fig_eq}.

The ``fast'' stability analysis (not presented here) shows that
small deviations $\theta(0)$ from an equilibrium solution, while all parameter values
are kept fixed, may result in two types of dynamics, depending
on the initial equilibrium $T_0$: fast increase or fast decrease of 
the initial deviation \citep{GC87}. 
The fast increase characterizes {\it unstable} equilibria: a small
deviation $\theta(0)$ from such an equilibrium $T_0$ 
forces the solution to go further and further away from the equilibrium.
In practice, such equilibria will not be observed, since
there are always small, random perturbations of the climate present 
in the system: just think of weather as representing such perturbations.

The fast decrease of the initial deviation $\theta(0)$ characterizes 
{\it stable} solutions;
only such equilibria can be observed in practice.
The two stable solution branches of \eqref{T1} are shown by solid
lines in Fig.~\ref{fig_eq}, while the unstable branch is shown by the dashed line.
The arrows show the direction in which the temperature will change
when drawn away from an equilibrium by external forces. This change,
whether away from or towards the nearest equilibrium, is fast
compared to the one that occurs along either solution branch \citep{MG76,MG94}.

\subsubsection{Bifurcation analysis}
\label{structural}

Given the choice of model parameters, the present climate 
state corresponds to the upper stable solution of Eq.~\eqref{T1}, at
$\mu=1$ (see Fig.~\ref{fig_eq}).
It lies quite close to the bifurcation point $(\mu,T)\approx(0.9,280~{\rm K})$, 
where the stable and unstable solutions merge.

The so-called {\it normal form} of this bifurcation is given by
the equation
\be
\dot{X}=\bar\mu-X^2,
\label{normal}
\ee 
where $X$ is a suitably normalized form of $T$, and $\bar\mu$
is a normalized form of $\mu$. 
Equation \eqref{normal} describes the dependence between $T$ 
and $\mu$ in a small neighborhood of the bifurcation point.
In particular, the stable equilibrium branch is described by
\[X=+\sqrt{\bar\mu};\]
this result has exactly the same form as the positive solution of  
Eq.~\eqref{TR}, given by our self-consistent analysis
of climate sensitivity in the presence of genuine nonlinearities,
cf.  Section \ref{sensitivity}.
Hence, the derivative ${\rm d}X/{\rm d}{\bar\mu}$, and thus 
${\rm d}T/{\rm d}\mu$, goes to infinity as 
the model approaches the bifurcation point;
this is exactly the situation discussed earlier in Section~\ref{sensitivity}.

It is important to realize that the parabolic form of temperature dependence
on insolation change is not an accident due to the particularly
simple form of EBMs. \citet{WM75} clearly showed, in a slightly
simplified GCM, that not only the mass-weighted
temperature of their total atmosphere, but also the area-weighted
temperatures of each of their five model levels, exhibits such a parabolic
dependence on fractional radiative input; see Fig. 5 in their paper. 
Moreover, these authors emphasize that
``As stated in the Introduction, it is not, however, reasonable to conclude 
that the present results are more reliable than the results from the 
one-dimensional studies mentioned above simply because our model 
treats the effect of transport explicitly rather than by parameterization.
[...] Nevertheless, it seems to be significant that both the one-dimensional 
and three-dimensional models yield qualitatively similar results in many respects."

In fact, rigorous mathematical results demonstrate that the {\em 
saddle-node bifurcation} whose normal form is given by Eq.~\eqref{normal}
occurs in several systems of nonlinear partial differential equations, such as 
the Navier-Stokes equations \citep{Const89,Temam97}, and not only
in ordinary differential equations, like Eqs.~\eqref{T} and \eqref{T1} above.
We emphasize, though, that this does not cause the temperature to increase
rapidly due to small changes in insolation: 
the presence of the bifurcation point will result in small, positive changes
of global temperature for slow, positive changes in $\mu$, while it
may throw the climate system into the deep-freeze state for slow,
negative changes in $\mu$.

\section{Discussion}
\label{discuss}

\subsection{How sensitive is climate?}

Making projections of climate change for the next decades and centuries, 
evaluating the human influence on future Earth temperatures, and making 
normative decisions about current and future anthropogenic impacts on 
climate are enormous tasks that require solid scientific expertise, as well as 
responsible moral reasoning. Well-founded approaches to handle
the moral aspects of the problem are still being debated [{\it e.g.}, \cite{HG08}].
It is that much more important to master existing tools for acquiring accurate
and reliable scientific evidence from the available data and models. 
Several of these tools come from the realm of nonlinear and complex dynamical 
systems \citep{Lorenz63,Smale67,RuelleTak,GC87,MG94,GCS08}. 

A straightforward analysis, carried out in Section~\ref{sensitivity} of this paper, 
shows that a proper treatment of the higher-order terms in a climate model
with nonlinear feedbacks does not reveal the exaggerated sensitivity to forcing 
that was used in RB07 to advocate intrinsic unpredictability of climate 
projections. 
We emphasize that the error in Roe and Baker's analysis is not
related to their choice of model formulation or of the model parameters nor 
to their interpretation of model results.
The problem is purely a matter of elementary calculus, and is due to inappropriate, 
and unnecessary, linearization of a nonlinear model. 

Our analysis complements, reinforces and goes beyond that of \cite{HDN+09}, 
who also showed that the claim of RB07 ``results from a mathematical artifact."
We notice simply that Hannart and colleagues did not even question the linear 
approximation framework of RB07 and still concluded that the claims of irreducibility
of the spread in the envelope of climate sensitivity are not supported by
the RB07 analysis.

To summarize, while the general {\it human} concern about climate sensitivity 
expressed by RB07 should be reasonably shared by many, their {\it scientific} 
conclusions do not follow from their model and its results, when correctly analyzed,
as done here in Section~\ref{sensitivity}. Nor are these conclusions supported 
by other models of greater detail and realism, when properly investigated.
Accordingly, conclusions about the likelihood of extreme warming resulting
from small changes in anthropogenic forcing can hardly be used to support 
political proposals [{\it e.g.}, \citet{AF08}] that claim to provide future 
directions for the climate-related sciences.

Still, this paper's analysis does not preclude in any sense the Earth's temperature 
from rising significantly in coming years. The methods illustrated here can only 
be used to study climate sensitivity in the vicinity of a given state; they cannot 
be applied to investigate climate evolution over tens of years, for example in 
response to large increases in greenhouse gases or to other major changes
in the forcing, whether natural or anthropogenic.
This latter problem requires global interdisciplinary efforts and, in particular,
the analysis of the entire hierarchy of climate models \citep{SD74}, from conceptual 
to intermediate to fully coupled GCMs \citep{GR00}. It also requires a much
more careful study of random effects than has been done heretofore \citep{GCS08}.

It seems to us that Roe and Baker's title question "Why Is Climate Sensitivity 
So Unpredictable?" still remains open.

\subsection{Where are the ``tipping points''?}
\label{cool}
The S-shaped diagram of Fig.~\ref{fig_eq} --- see also Fig.~10.6 in \citet{GC87}
and Fig. 4 in \citet{MG94} --- was  used here to show the 
smoothness and boundedness of temperature changes as a function of 
insolation changes, away from a saddle-node bifurcation, like that of 
Eq.~\eqref{TR} in Section~\ref{sensitivity} or of Eq.~\eqref{normal} in Section
~\ref{structural}. 

This S-shaped curve nevertheless reveals the existence of sensitive 
dependence of Earth's temperature on insolation changes, or on other
changes in Earth's net radiation budget, such as may be caused by
increasing levels of greenhouse gases, on the one hand, or of aerosols,
on the other. This sensitive dependence is quite
different from the one advocated by RB07. 
Namely, if the parameter $\mu$ were to slightly decrease
--- rather than increase, as it seems to have done since the mid-1970s,
in the sense described in the last paragraph of Section~\ref{model} ---
then the climate system would be pushed past the bifurcation point 
at $\mu\approx0.9$. 
The only way for the global temperature to go would be down, all the
way to a deep-freeze Earth, with much lower temperatures than those 
of recent, Quaternary ice ages.

It has become common in recent discourse about potentially irreversible 
climate change to talk about ``tipping points"; {\it e.g.}, \cite{Lenton+08}.
The term was originally introduced into the social sciences by \cite{Glad00}
to denote a point at which a previously rare phenomenon becomes dramatically 
more common. In the physical sciences, it has been identified with a shift from one
stable equilibrium to another one, {\it i.e.}, to a saddle-node bifurcation,
as seen in Fig.~\ref{fig_eq} here and explained in Section~\ref{structural} above.

In the EBM context of Fig.~\ref{fig_eq}, it would require an enormous, 
almost twofold increase in the insolation in order for a deep-freeze--type 
equilibrium to reach the bifurcation point at $\mu\approx 1.85$ and 
jump from there to $T\approx 350$ K, a temperature that
sounds equally unpleasant. Within the broader context of the recent
debates on how to exit a snowball-Earth state, very large, and possibly
implausible increases in ${\rm CO}_2$ levels would be required
\citep{Pie04}. 

Indeed, the likelihood to actually reach the tipping point to the left of the 
current climate in Fig.~\ref{fig_eq} seems to be quite small. Mechanisms 
for entering a snowball-Earth climate have been 
recently studied with a number of fairly realistic climate models 
\citep{Hyde+00,D+04,PJ04}.
Both modeling and independent geological evidence suggest that Earth's 
climate can sustain significant fluctuations of the solar radiative input, 
and hence of global temperature, without entering the snowball Earth,
and evidence for Earth ever having been in such a state is still controversial.   

Nevertheless, the existence of the upper-left tipping point shown 
in Fig.~\ref{fig_eq} is confirmed 
by numerous model studies, including GCMs, and we have already cited some
evidence also for the lower-right tipping point in the figure.
Several hypothetical tipping points on the ``warm" side have been identified by 
\citet{Lenton+08} and references therein, among many others. But only
few of these have been studied with the same degree of mathematical 
and physical detail as the ones of Fig.~\ref{fig_eq} here. One worthwhile
example is that of the oceans' buoyancy-driven, or thermohaline,
circulation \citep{Stommel61,Bryan,Quon92,Thual92,DG05}.

Accordingly, humankind must be careful --- in pursuing its recent
interest in geoengineering \citep{Cru06,Mac06} --- to stay a course
that runs between tipping points on the warm, as well as on the ``cold" side 
of our current climate. In any case, the existence, position and properties of such
tipping points need to be established by physically careful and
mathematically rigorous studies. The ``margins of maneuver'' seem 
reasonably wide, at least on the time scale of tens to hundreds of years, 
but this does not eliminate the possibility to eventually reach one such 
tipping point, and thus we are led directly to the next question.

\subsection{How close are we to a cold tipping point?}

Let us assume for the moment that the dangers of further warming will
lead humanity to actually stop, and possibly reverse, the current trend
of an increasingly positive net radiation balance. 
Given, on the other hand, the dangers of a snowball Earth,
one might want to estimate then the closeness of the climate system 
to the top-left bifurcation point in Fig.~\ref{fig_eq} here.

The GCM simulations of \citet{WM75} (see again their Fig. 5) suggest
that this point might lie no farther than $5\%$ below the current value of 
the solar constant. At the same time, the Sun has been much fainter 
4 Gyr ago (by approximatyely 25--30\%) than today, without the Earth
ending up in a deep freeze, except possibly much later.
So how close are we to this tipping point?

\begin{figure}[p]
\centering\includegraphics[width=8.3cm]{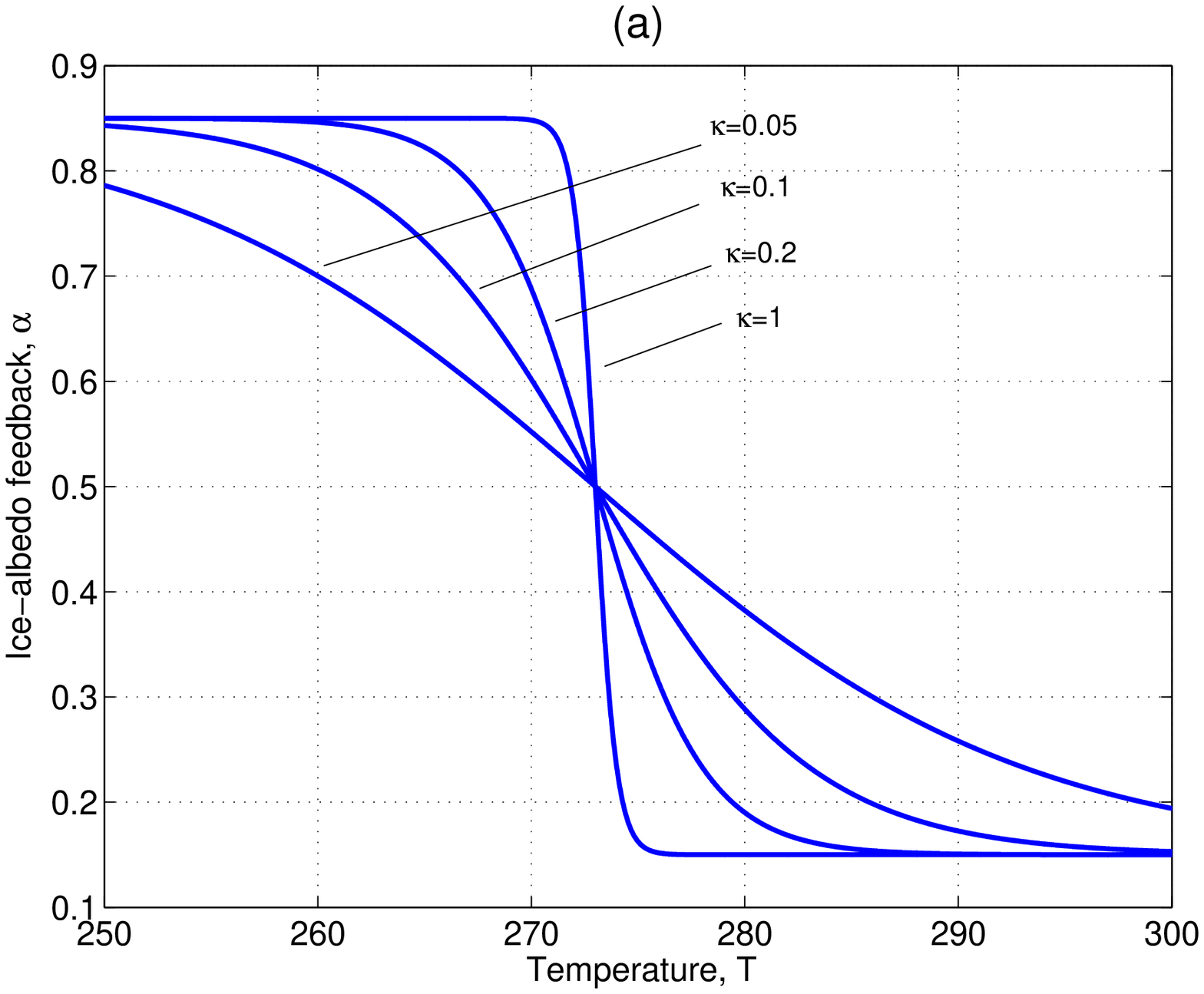}
\centering\includegraphics[width=8.3cm]{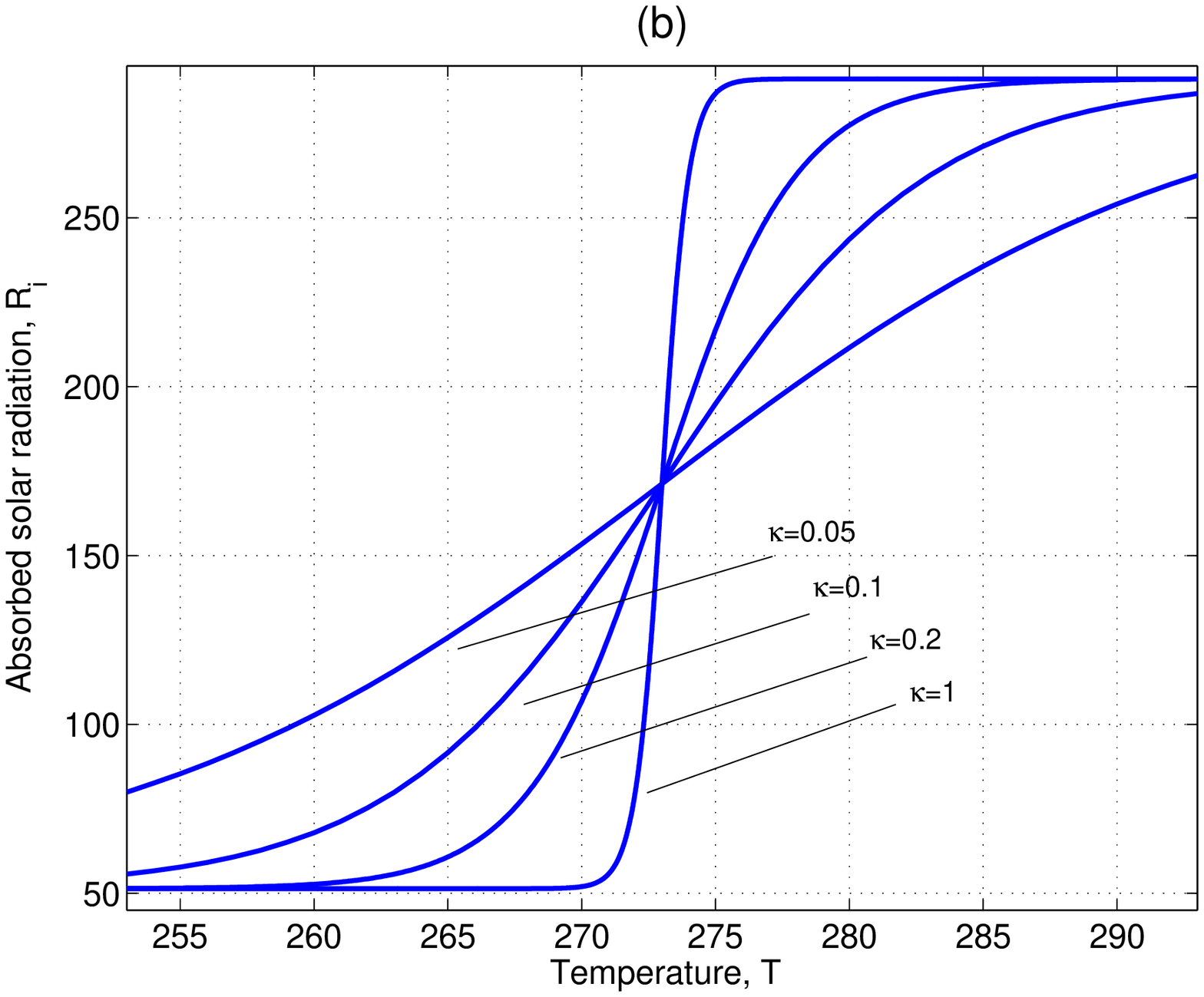}
\caption{Dependence of the absorbed incoming radiation 
$R_{\rm i}$ on the steepness parameter $\kappa$:
(a) ice-albedo feedback $\alpha=\alpha(T;\kappa)$, and (b)
absorbed radiation $R_{\rm i}=R_{\rm i}(T;\kappa)$, 
for different values of $\kappa$; 
see Eqs.~\eqref{Ri}~and~\eqref{albedo}.} 
\label{fig_kappa}
\end{figure}

\begin{figure}[p]
\centering\includegraphics[width=8.3cm]{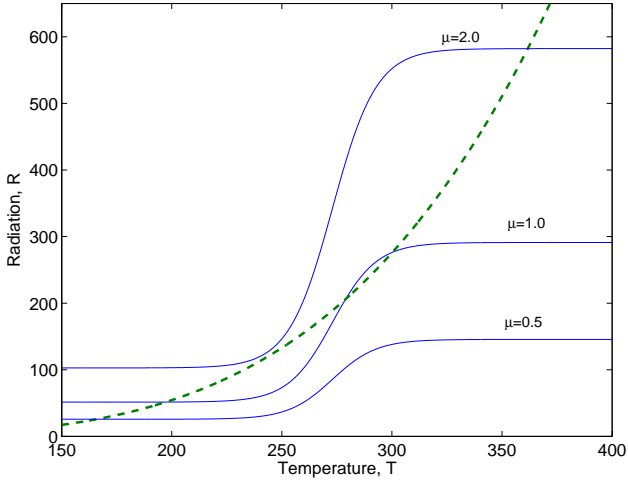}
\caption{Outgoing radiation $R_{\rm o}$ (red, dashed line) and
absorbed incoming solar radiation $R_{\rm i}$ (blue, solid lines)
for our 0-D energy-balance model (EBM), governed by Eq.~\eqref{T1}.
The absorbed radiation is shown for $\mu=0.5,1$ and 2.0 (from
bottom to top), while $\kappa=0.05$.
Notice the existence of one or three intersection points between the 
$R_{\rm o}$ curve and one of the $R_{\rm i}$ curves, depending 
on the value of $\mu$; these points correspond to the equilibrium
solutions of \eqref{T1}, {\it i.e.} to steady-state climates.}
\label{fig_TR}
\end{figure}

\begin{figure}[p]
\centering\includegraphics[width=8.3cm]{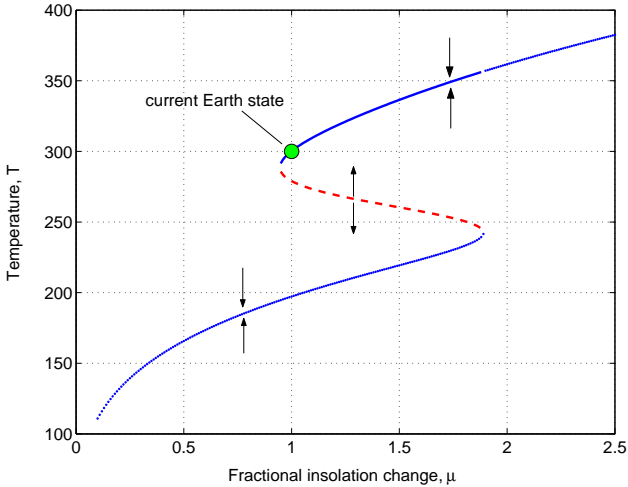}
\caption{Equilibrium solutions of the EBM \eqref{T1}
depending on the fractional change $\mu$ in insolation.
Notice the existence of two stable (blue, solid lines) and one unstable 
(red, dashed line) solution branches.
The arrows show the direction in which the global temperature will change
after being displaced from a nearby equilibrium state by external forces.
The current Earth state corresponds to the upper stable solution
at $\mu=1$; in this figure $\kappa=0.05$.} 
\label{fig_eq}
\end{figure}

\begin{figure}[p]
\centering\includegraphics[width=8.3cm]{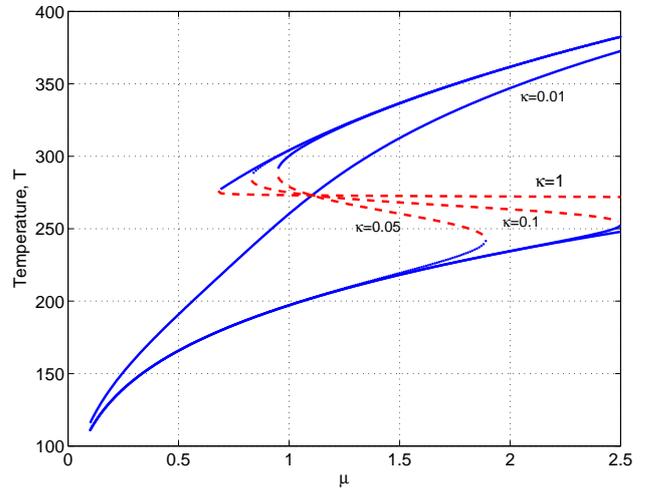}
\caption{Equilibrium solutions of the EBM \eqref{T1}
for ice-albedo feedback functions $\alpha=\alpha(T;\kappa)$ 
given by Eq.~\eqref {albedo} and corresponding to
the steepness parameter values $\kappa=0.01, 0.05, 0.1$ and 1.
Notice the deformation of the stable (blue, solid lines) and 
unstable (red, dashed lines) solution branches: the ``cold" tipping 
point moves to the left, away from $\mu=1.0$, as $\kappa$ increases;
compare Fig.~\ref{fig_kappa}a.}
\label{fig_eqs}
\end{figure}

Figure~\ref{fig_eqs} here shows stable and unstable equilibrium solutions
for different profiles of the ice-albedo feedback, 
$\alpha=\alpha(T;\kappa)$; this profile is determined by the 
value of the steepness parameter $\kappa$ (cf. Fig.~\ref{fig_kappa}a).
The figure suggests that the steeper the ramp of the ice-albedo feedback 
function, {\it i.e.} the larger $\kappa$, the further away the bifurcation might lie.
In fact, for a very smooth dependence of the albedo on temperature,
{\it i.e.} for a very small $\kappa$, there is no bifurcation at all (not shown): 
very small values of $\kappa$ produce only a single-valued, smoothly 
increasing, stable equilibrium solution to \eqref{T1} for any value of $\mu$.

It seems worthwhile to carry out systematic bifurcation studies 
with atmospheric, oceanic and coupled GCMs to examine this question more
carefully, for warm tipping points, as well as for cold ones. Such studies are made 
possible by current computing capabilities, along with well-developed 
methods of numerical bifurcation theory \citep{DG05,SDG09}.
This approach holds some promise in evaluating the distance of the 
current climate state from either a catastrophic warming or a catastrophic 
cooling.

{\bf Acknowledgements.}
We thank two anonymous referees for their constructive comments that
helped improve the paper.
This study was supported by U.S. Department of Energy grants 
DE-FG02-07ER64439 and DE-FG02-07ER64440 from 
its Climate Modeling Programs.

\newpage

\end{document}